\documentclass[twocolumn]{aastex63}

\graphicspath{{./}{figures/}}
\usepackage{amsmath}
\begin{document}

\title{High Energy Neutrinos from Choked Gamma-Ray Bursts in AGN Accretion Disks}

\author[0000-0002-9195-4904]{Jin-Ping Zhu}
\affil{Department of Astronomy, School of Physics, Peking University, Beijing 100871, China; \url{zhujp@pku.edu.cn}}

\author[0000-0003-4976-4098]{Kai Wang}
\affiliation{Department of Astronomy, School of Physics, Huazhong University of Science and Technology, Wuhan 430074, China}

\author[0000-0002-9725-2524]{Bing Zhang}
\affiliation{Department of Physics and Astronomy, University of Nevada, Las Vegas, NV 89154, USA; \url{zhang@physics.unlv.edu}}

\author[0000-0001-6374-8313]{Yuan-Pei Yang}
\affiliation{South-Western Institute for Astronomy Research, Yunnan University, Kunming, Yunnan, People’s Republic of China}

\author[0000-0002-1067-1911]{Yun-Wei Yu}
\affiliation{Institute of Astrophysics, Central China Normal University, Wuhan 430079, China}

\author[0000-0002-3100-6558]{He Gao}
\affiliation{Department of Astronomy, Beijing Normal University, Beijing 100875, China}

\begin{abstract}
Both long-duration gamma-ray bursts (LGRBs) from core collapse of massive stars and short-duration GRBs (SGRBs) from mergers of binary neutron star (BNS) or neutron star--black hole (NSBH) are expected to occur in the accretion disk of active galactic nuclei (AGNs). We show that GRB jets embedded in the migration traps of AGN disks are promised to be choked by the dense disk material. Efficient shock acceleration of cosmic rays at the reverse shock is expected, and high-energy neutrinos would be produced. We find that these sources can effectively produce detectable TeV--PeV neutrinos through $p\gamma$ interactions. From a choked LGRB jet with isotropic equivalent energy of $10^{53}\,{\rm erg}$ at $100\,{\rm Mpc}$, one expects $\sim2\,(7)$ neutrino events detectable by IceCube (IceCube-Gen2). The contribution from choked LGRBs to the observed diffuse neutrino background depends on the unknown local event rate density of these GRBs in AGN disks. For example, if the local event rate density of choked LGRBs in AGN disk is $\sim5\%$ that of low-luminosity GRBs $(\sim10\,{\rm Gpc}^{-3}\,{\rm yr}^{-1})$, the neutrinos from these events would contribute to  $\sim10\%$ of the observed diffuse neutrino background. Choked SGRBs in AGN disks are potential sources for future joint electromagnetic, neutrino, and gravitational wave multi-messenger observations.
\end{abstract}


\section{Introduction} \label{sec:intro}

Massive stars can be born in situ in an AGN accretion disk or be captured from the nuclear star cluster around the AGN \cite[e.g.,][]{artymowicz1993,collin1999,wang2011,fabj2020,Cantiello2020,dittmann2021}. When they die, these massive stars will make supernovae (SNe) and leave behind NS and BH remnants inside the disk. Most of them might have extreme high spins and easily make LGRBs \citep{jermyn2021}. Such embedded massive stars and stellar remnants would migrate inwards the trapping orbits within the disk \cite[e.g.,][]{bellovary2016,tagawa2020}. Abundant compact objects within the orbit would likely collide and merge \citep{cheng1999,mckernan2020}, which are the promising astrophysical gravitational wave (GW) sources for LIGO \citep{abbott2009}. A possible SN \citep{assef2018} and a candidate binary black hole merger induced electromagnetic transient \citep{graham2020} have been reported recently in association with AGN disks. 

GRBs, both long-duration ones associated with core collapse of massive stars \citep{woosley1993,macfadyen1999,zhang2003,zhang2004,galama1998,hjorth2003,stanek2003} and short-duration ones associated with neutron star mergers \citep{paczynski1986,paczynski1991,eichler1989,narayan1992,abbott2017a,abbott2017b,abbott2017c} have been suggested as sources of astrophysical high-energy cosmic rays and neutrinos \citep{waxman1997}. 
On the other hand, non-detection of GRB neutrino signals \citep{abbasi2012,aartsen2015b}, likely related to a large emission radius from the central engine \citep{zhangkumar2013}, suggested GRB-associated neutrinos can only account for at most $\lesssim1\%$ of the diffuse neutrino fluence \citep{murase2008,wang2009}. The so-called low-luminosity GRBs are more abundant than successful ones \citep{liang2007,sun2015} and can give significant contribution to the neutrino background \citep{murase2006,gupta2007}. One possibility is that they originate from massive stars that launched jets that are choked in the stellar envelope \citep{bromberg2011l,nakar2015}. Such choked jets from the death of massive stars or even from neutron star mergers could be more promising sites for efficient neutrino production, which may contribute to a considerable fraction of the diffuse neutrino background \citep{murase2013,senno2016,xiao2014,kimura2018,fasano2021}. 

The presence of massive stars and compact  binaries in AGN disks indicates that both LGRBs and SGRBs could possibly occur in such a high density environment. Very recently, \cite{zhu2021,perna2021} suggested that GRBs jet in the AGN disks are likely choked\footnote{\cite{kimura2021} recently presented that compact binaries can accrete, produce radiation-driven outflows, and create cavities in the AGN disks before the merger, so that aligned SGRB jets could  successfully break out from the AGN disks.}, instead of making  and result in observable signals from optical to $\gamma$-ray. In this work, we consider GRB jets choked inside the AGN disks as hidden sources of high energy cosmic neutrinos, which can ease the tension between the diffuse extragalactic gamma-ray background and the diffuse background of TeV--PeV neutrinos.

\section{Jet Dynamics in AGN Accretion Disks}

SNe and neutron star mergers are expected to occur in the migration traps \citep{bellovary2016} plausibly located at $a\sim 10^3 r_{\rm g}$, where $r_{\rm g} \equiv GM_{\rm SMBH}/c^2$, $G$ is the gravitational constant, $M_{\rm SMBH}$ is the AGN supermassive BH mass, and $c$ is the speed of light. For a gas-pressure-dominated disk, the disk density is $\rho(z) = \rho_0\exp(-z/H)$, where $\rho_0$ is the midplane density, $z$ is the vertical distance, and $H= 1.5\times10^{14}\,M_{\rm SMBH,8\odot}\left( \frac{H/a}{0.01} \right)\left( \frac{a}{10^3 r_{\rm g}}\right)$ is the disk height with the disk aspect ratio $H / a\sim[10^{-3} , 0.1]$ \citep{goodman2004,thompson2005}. Near the migration traps, the midplane density is almost $\rho_0\sim O(10^{-10})\,{\rm g}\,{\rm cm}^{-3}$ and the disk height is $H\sim O(10^{14})\,{\rm cm}$. For a disk with an exponentially decaying density profile, the density $\rho\approx\rho_0$ for $z<H$ while $\rho$ decreases rapidly for $z>H$. Therefore, for simplicity we approximately adopt a uniform density profile for the AGN disk for $z<H$. Hereafter, the convention $Q_x = Q / 10^x$ is adopted in cgs units.

\begin{figure}[tpb]
    \centering
    \includegraphics[width = 1\linewidth , trim = 0 100 0 55, clip]{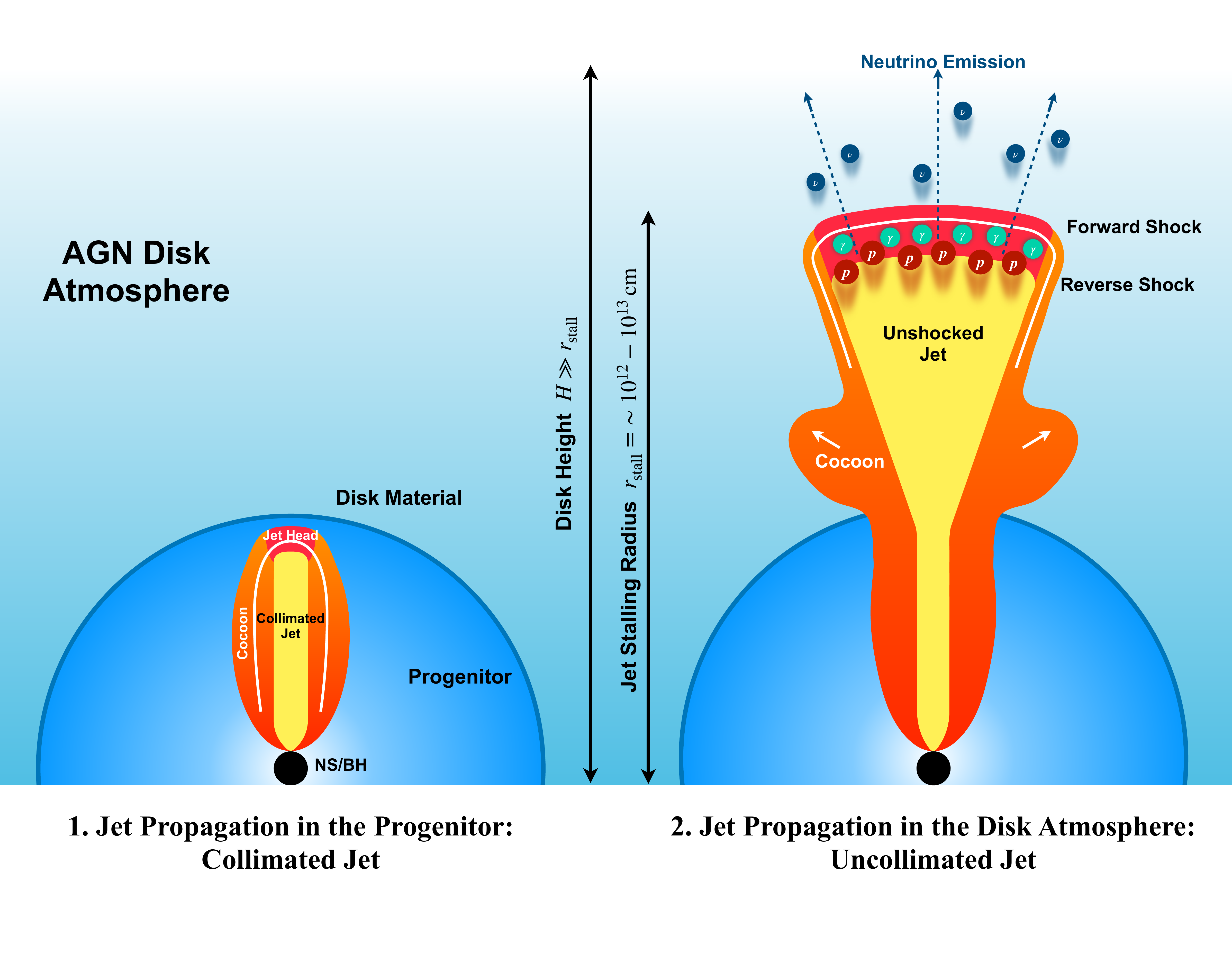}
    \caption{Schematic picture of jet propagation in the progenitor and in the AGN disk.}
    \label{fig:Cartoon}
\end{figure}

Figure\,\ref{fig:Cartoon} illustrates the physical processes for a jet traveling through the progenitor star and the AGN accretion disk. When a jet initially propagates inside the progenitor star, the collision between the jet and the stellar gas medium leads to the formation of a forward shock sweeping into the medium and a reverse shock entering the jet material \cite[e.g.,][]{matzner2003,bromberg2011,yu2020}. Such a structure is known as the jet head. The hot material that enters the head flows sideways and produces a powerful cocoon to drive a collimation shock into the jet material. The jet is collimated inside the star and gets accelerated to a relativistic velocity after it breaks out from the progenitor star. 

After the jet entering into the AGN disk, the jet head velocity is given by \citep{matzner2003}: $\beta_{\rm h} = \beta_{\rm j}/ (1 + \tilde{L}^{-1/2})$, where $\beta_{\rm j}\simeq 1$ and $\tilde{L} \equiv L_{\rm j}/\pi r_{\rm j}^2\rho_0c^3\approx (10^{14}\,{\rm cm}/r_{\rm j})^2L_{{\rm j},50}\rho_{0,-10}^{-1}$ is the critical parameter that determines the evolution of the jet \citep{bromberg2011}, $L_{\rm j}$ is the jet luminosity, and $r_{\rm j}$ is the radius of the jet from the central engine. Since this jet radius should be $r_{\rm j} \ll 10^{14}\,{\rm cm}$, one gets $\tilde{L}>\theta_0^{-4/3}\gg$1, where $\theta_0 \approx 0.2$. In this case, the jet head would travel with a relativistic speed. The cocoon pressure is too weak to affect the geometry of the jet so that the jet is uncollimated (i.e., $\theta_{\rm j} \approx \theta_0$)\citep{bromberg2011}. Therefore, the critical parameter can be expressed as $\tilde{L} = L_{\rm iso} / 2\pi r_{\rm h}^2\rho_0c^3$, where $L_{\rm iso} \approx 2L_{\rm j} / \theta_0^2$ is the isotropic-equivalent one-side jet luminosity and $r_{\rm h}\approx r_{\rm j} / \theta_0$ is the  distance between the jet head and the central engine. The Lorentz factor of the jet head is given by $\Gamma_{\rm h} \approx \tilde{L}^{1/4} / \sqrt{2}$. The internal energy and density evolution of the forward shock and reverse shock can be described by the shock jump conditions \citep{blandford1976,sari1995,zhang2018}:
\begin{equation}
\begin{split}
\label{eq:ShockJumpCondition}
    e'_{\rm f}/n'_{\rm f}m_pc^2 &= \Gamma_{\rm h} - 1,~n'_{\rm f}/n_{\rm a} = (\hat{\gamma}_{\rm 1}\Gamma_{\rm h} + 1)/(\hat{\gamma}_{\rm 1} - 1)=4\Gamma_{\rm h},\\
    e'_{\rm r}/n'_{\rm r}m_pc^2 &= \bar{\Gamma}_{\rm h} - 1,~n'_{\rm r}/n''_{\rm j} = (\hat{\gamma}_{\rm 2}\bar{\Gamma}_{\rm h} + 1)/(\hat{\gamma}_{\rm 2} - 1)=4\bar{\Gamma}_{\rm h},
\end{split}
\end{equation}
where $\hat{\gamma}_1 = (4\Gamma_{\rm h} + 1)/3\Gamma_{\rm h}$, $\hat{\gamma}_2 = (4\bar{\Gamma}_{\rm h} + 1)/3\bar{\Gamma}_{\rm h}$, the subscripts ``a'', ``f'', ``r'' and ``j'' represent regions of the unshocked AGN material, the jet head's forward shock, the jet head's reverse shock, and the unshocked jet, and $\bar{\Gamma}_{\rm h} = \Gamma_{\rm j}\Gamma_{\rm h}(1 - \beta_{\rm j}\beta_{\rm h})\approx \Gamma_{\rm j}\tilde{L}^{-1/4}/\sqrt{2}$ is the Lorentz factor of the unshocked jet measured in the jet head frame. Note we distinguish among three reference frames: $Q$ for the AGN rest frame, $Q'$ for the jet head comoving frame, and $Q''$ for the jet comoving frame. 

We assume that a luminous jet can easily break out from the progenitor star, which has a similar parameter distribution as classical GRBs. Observationally, the average duration for LGRBs is $t_{\rm j} \approx 10^{1.5}\,{\rm s}$ \citep{koiuveliotou1993,horvath2002,zhang2018} while the median isotropic energy release is $E_{\rm iso}\approx10^{53}\,{\rm erg}$ for LGRB jets (the isotropic equivalent luminosity $L_{\rm iso} = E_{\rm iso}/t_{\rm j}$)\citep{kumar2015}. The jet could be choked in the AGN disk when the central engine quenches so that the jet head radius at $t_{\rm j}$ can be defined as the stalling radius, i.e., $r_{\rm stall} = r_{\rm h}(t_{\rm j}) \approx 2\Gamma_{\rm h}^2ct_{\rm j} = 2.0\times10^{13}\,E_{{\rm iso},53}^{1/4}t_{{\rm j},1.5}^{1/4}\rho_{0,-10}^{-1/4}\,{\rm cm}$, with $\Gamma_{\rm h} = 2.7\,E_{{\rm iso},53}^{1/8}t_{{\rm j},1.5}^{-3/8}\rho_{0,-10}^{-1/8}$. For SGRB jets, the average duration and the median isotropic energy are $t_{\rm j}\approx10^{-0.1}\,{\rm s}$ and $E_{\rm iso}\approx10^{51}\,{\rm erg}$ \citep{fong2015}, respectively. The stalling radius for SGRB jet is $r_{\rm stall} = 2.5\times10^{12}\,E_{{\rm iso},51}^{1/4}t_{{\rm j},-0.1}^{1/4}/\rho_{0,-10}^{-1/4}\,{\rm cm}$ and the jet head Lorentz factor when it chokes is $\Gamma_{\rm h} = 6.1\,E_{{\rm iso},51}^{1/8}t_{{\rm j},-0.1}^{-3/8}\rho_{0,-10}^{-1/8}$. Because $r_{\rm stall} \ll H$, both LGRB and SGRB jets can be easily choked in an AGN accretion disk.

\begin{figure*}[tpb]
    \centering
    \includegraphics[width = 0.49\linewidth , trim = 50 30 70 57, clip]{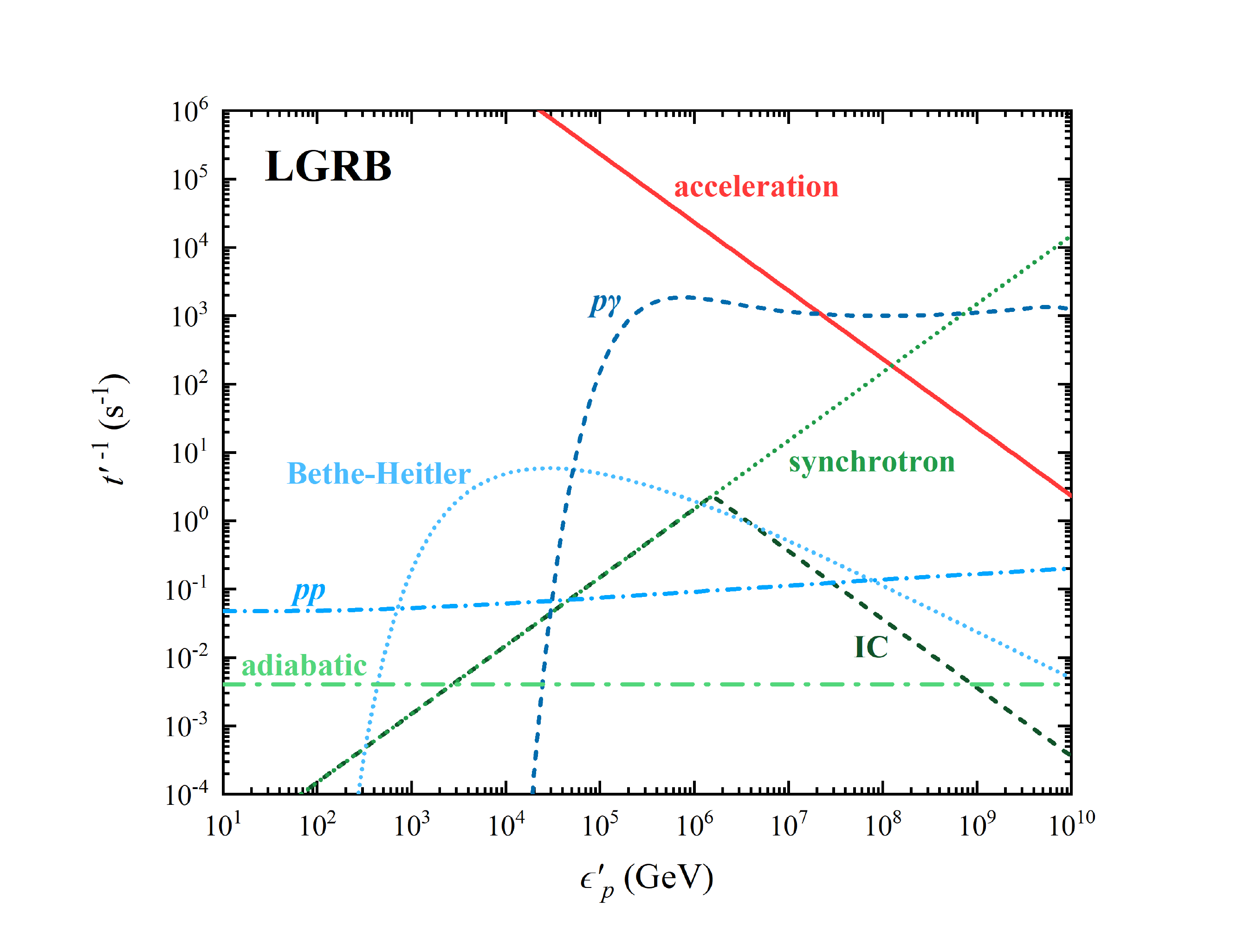}
    \includegraphics[width = 0.49\linewidth , trim = 50 30 70 57, clip]{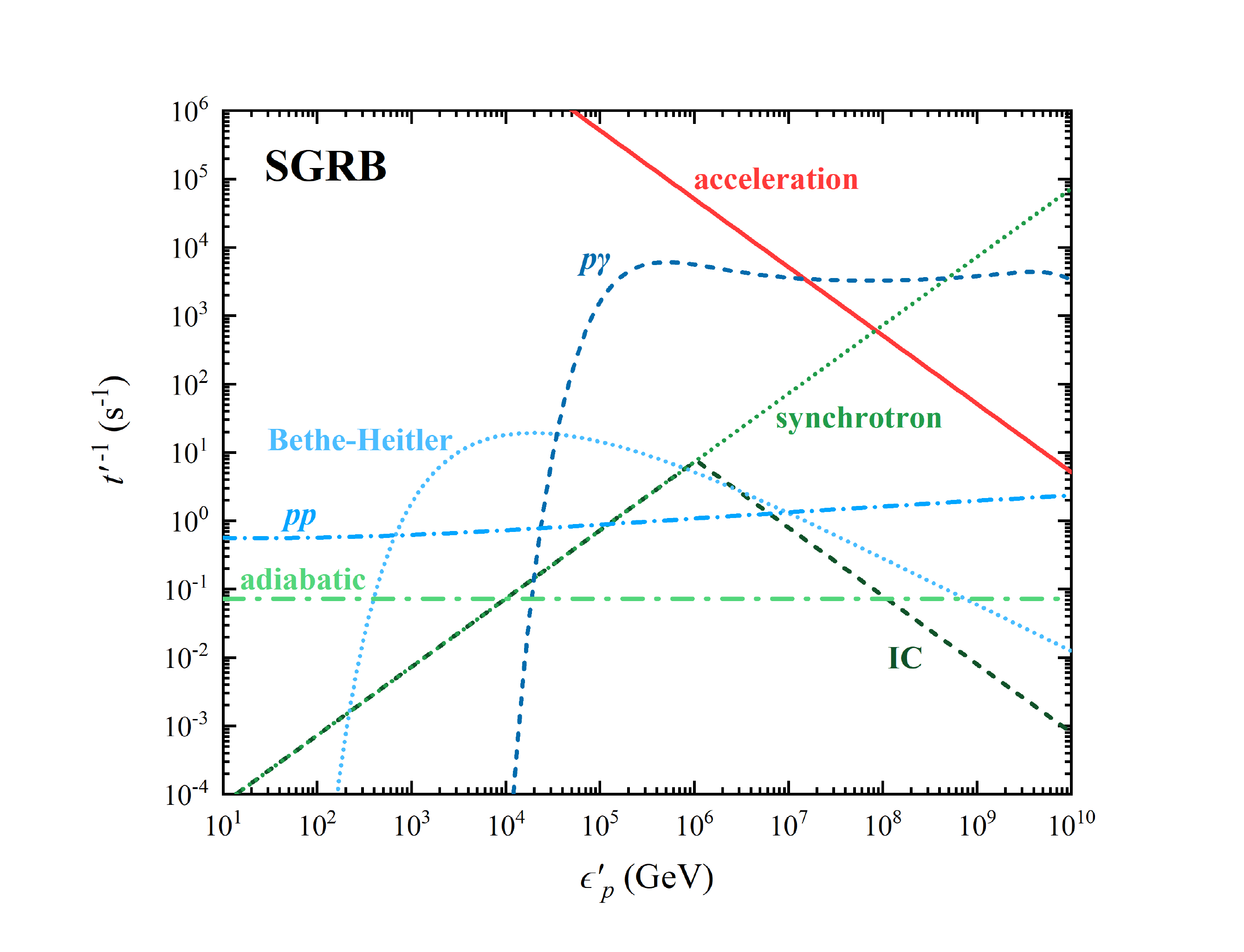}
    \caption{Inverse of proton acceleration and cooling timescales as a function of proton energy in the jet head frame for a classical LGRB (left panel) and SGRB (right panel). Acceleration (red solid), photomeson production ($p\gamma$, blue dashed), Bethe-Heitler pair production (blue dotted), hadronuclear scattering ($pp$, blue dashed-dotted), inverse-Compton (IC, green dashed), synchrotron radiation (green dotted), and adiabatic cooling (green dashed-dotted) processes are considered.}
    \label{fig:Time}
\end{figure*}

\section{Neutrino Production}

We assume that the Fermi acceleration operates and accelerated protons have a power-law distribution in energy: $dn_p/d\epsilon_p\propto\epsilon^{-s}_p$ with $s = 2$ \citep{achterberg2001,keshet2005}. Here, the thermal photons from the jet head are treated as the only background photon field for hadronic interactions since the number densities of other types of radiation (e.g., the classical keV--MeV emission of GRBs) are typically much lower \citep{senno2016}. Based on Eq.\,(\ref{eq:ShockJumpCondition}), the internal energy and proton energy density of the jet head can be expressed as $e'_{\rm r} = (\bar{\Gamma}_{\rm h} - 1)n'_pm_pc^2$ and $n'_p = 4\bar{\Gamma}_{\rm h}n''_{\rm j}$, where $m_p$ is the proton mass and $n''_{\rm j} = L_{\rm iso}/4\pi\Gamma_{\rm j}^2r_{\rm stall}^2m_pc^3$ is the jet density. The photon temperature of the jet head is $k_{\rm B}T'_{\rm r} = (15\hbar^3c^3 \varepsilon_ee'_{\rm r}/\pi^2)^{1/4}$, where $\varepsilon_{e}\approx 0.1$ is the electron energy fraction. For classical parameters of LGRB (SGRB) jets, $k_{\rm B}T'_{\rm r} \approx 0.21 (0.32)\,{\rm keV}$. One can obtain the average thermal photon energy $\epsilon'_\gamma = 2.7k_{\rm B}T'_{\rm r}$ and the average thermal photon density $n'_\gamma = \varepsilon_ee'_{\rm r}/\epsilon'_\gamma$.

For the reverse shock, efficient Fermi acceleration can occur only if the radiation constraint \citep{murase2013} is satisfied, i.e., $\tau_{\rm T} = n''_{\rm j}\sigma_{\rm T}r_{\rm stall}/\Gamma_{\rm j}\lesssim {\rm min}[{1,0.1C^{-1}\bar{\Gamma}_{\rm h}}]$, where $C = 1 + 2\ln\bar{\Gamma}_{\rm h}^2$. By considering $\Gamma_{\rm j} = 300$ (500) for LGRB (SGRB) jets, we get $\tau_{\rm T} = 6.8\times10^{-3}\,E_{\rm iso,53}^{3/4}t_{\rm j,1.5}^{-5/4}\rho_{0,-10}^{1/4}\Gamma_{\rm j,2.5}^{-3}\ll0.32C^{-1}_{1.2}\bar{\Gamma}_{\rm h,1.7}\ll1$ ($\tau_{\rm T} = 4.6\times10^{-3}\,E_{\rm iso,51}^{3/4}t_{\rm j,-0.1}^{-5/4}\rho_{0,-10}^{1/4}\Gamma_{\rm j,2.7}^{-3}\ll0.26C^{-1}_{1.2}\bar{\Gamma}_{\rm h,1.6}\ll1$), which means that Fermi acceleration is always effective. Note that different from \cite{senno2016} whose protons are from the internal shocks, the interacted protons in our calculations are accelerated from the reverse shock.

With the assumption of perfectly efficient acceleration, the acceleration timescale is given by $t'_{p,{\rm acc}} = \epsilon'_p/(eB'c)$, where the jet head comoving magnetic field strength is $B' = \sqrt{8\pi\varepsilon_Be'_r} \approx (\bar{\Gamma}_{\rm h}/\Gamma_{\rm j})(8\varepsilon_BL_{\rm iso}/r_{\rm stall}^2c)^{1/2}$ and the magnetic field energy fraction is assumed as $\varepsilon_B = 0.1$.

A high energy proton loses its energy through radiative, hadronic, and adiabatic processes. The radiative cooling mechanisms contain synchrotron radiation with cooling timescale 
\begin{equation}
\label{eq:syn}
     t'_{p,\rm syn} = \frac{6\pi m_p^4c^3}{\sigma_{\rm T}m_e^2B'^2\epsilon_p'},
\end{equation}
and inverse-Compton scattering with coolng timescale 
\begin{equation}
\label{eq:IC}
t'_{p,\rm IC} = \left\{\begin{matrix}
\frac{3m_p^4c^3}{4\sigma_{\rm T}m_e^2{n}'_\gamma{\epsilon}'_\gamma\epsilon_p'},~&{\epsilon}'_\gamma\epsilon_p'<m_p^2c^4,\\
\frac{3\epsilon_\gamma'\epsilon_p'}{4\sigma_{\rm T}m_e^2c^5{n}'_\gamma},~&{\epsilon}'_\gamma\epsilon_p'>m_p^2c^4.
\end{matrix}\right.
\end{equation}

The hadronic cooling mechanisms mainly include the inelastic hadronuclear scattering ($pp$), the Bethe-Heitler pair production ($p\gamma\to pe^+e^-$), and the photomeson production ($p\gamma$). High energy neutrinos are expected to be produced via $pp$ and $p\gamma$ processes. The cooling timescale of $pp$ scattering is given by $t'_{p,pp} = 1 / c\sigma_{pp}n'_p\kappa_{pp}$, where the inelasticity is set as $\kappa_{pp}\simeq0.5$ and the cross section $\sigma_{pp}$ is obtained from \cite{kelner2006}. The energy loss rate of $p\gamma$ production is calculated by the formula given in \cite{stecker1968,murase2007}, i.e.
\begin{equation}
\label{eq:pgamma}
    t'^{-1}_{p,p\gamma} = \frac{c}{2\gamma_p}\int^\infty_{\bar{\epsilon}_{\rm th}}d\bar{\epsilon}\sigma_{p\gamma}(\bar{\epsilon})\kappa_{p\gamma}(\bar{\epsilon})\bar{\epsilon}\int^\infty_{\bar{\epsilon}/2\gamma_p}d\epsilon\epsilon^{-2}\frac{dn}{d\epsilon},
\end{equation}
where $\bar{\epsilon}$ represents the photon energy in the rest frame of the proton, $\bar{\epsilon}_{\rm th} \simeq 145\,{\rm MeV}$ is the threshold energy, $\gamma_p = \epsilon'_p/m_pc^2$, and $dn/d\epsilon$ is the photon number density in the energy range of $\epsilon$ to $\epsilon + d\epsilon$. The inelasticity $\kappa_{p\gamma}$ and the cross section $\sigma_{p\gamma}$ are taken from \cite{stecker1968, patrignani2016}. The energy loss rate of Bethe-Heitler process $t'^{-1}_{\rm BH}$ can be also estimated based on Eq.\,(\ref{eq:pgamma}) by using $\kappa_{\rm BH}$ and $\sigma_{\rm BH}$ instead of $\kappa_{p\gamma}$ and $\sigma_{p\gamma}$. $\kappa_{\rm BH}$, $\sigma_{\rm BH}$ and $\bar{\epsilon}_{\rm th}$ for Bethe-Heitler process are adopted from \cite{chodorowski1992}.

Finally, the timescale that protons to lose energy due to adiabatic cooling is $t'_{p,{\rm ada}} = r_{\rm stall} / c\Gamma_{\rm h}$. We present the acceleration and cooling timescales of a choked LGRB and SGRB jet in an AGN disk in Figure\,\ref{fig:Time}. For both cases, $pp$ scattering would dominate the cooling process for low-energy protons. Bethe-Heitler process leads and suppresses neutrino production if the energy of protons falls within the range of $0.5\,{\rm TeV}\lesssim\epsilon'_p\lesssim20\,{\rm TeV}$. At higher energies, the dominant cooling mechanism for protons is $p\gamma$ interaction, which also limits the maximum proton energy to $\epsilon'_{p,{\rm max}} \sim 10\,{\rm PeV}$. 

Pions and kaons created through $pp$ and $p\gamma$ processes decay into muons and muon neutrinos. Pions and kaons are subject to hadronic scattering, $t'_{\{\pi,K\},{\rm had}} = 1/ c\sigma_{\rm h}n'_p\kappa_{\rm h}$, where $\sigma_{\rm h} \approx 5\times10^{-26}\,{\rm cm}^2$ and $\kappa_{\rm h}\approx 0.8$ \citep{olive2014}. The intermediate muons then decay to muon neutrinos, electron neutrinos and electrons. Similar to protons, pions, kaons, and muons also experience radiative processes and adiabatic cooling. One can calculate the synchrotron and IC cooling timescales of pions, kaons, and muons by Eq.\,(\ref{eq:syn}) and Eq.\,(\ref{eq:IC}) with $\epsilon'_{p}\to \epsilon'_{i}$ and $m_p\to m_i$, where $i = \pi,\,K,\,\mu_\pi,\,$ and $\mu_K$ are the parent particles for the neutrinos. The energy fractions from a proton to intermediate particles are calculated according to \cite{denton2018}. By comparing these cooling timescales with the decay timescales of intermediate particles, i.e., $t_{i,{\rm dec}} = \gamma_i\tau_i$ (where $\gamma_i = \epsilon'_i/m_ic^2$ and $\tau_i$ are the Lorentz factor and the rest frame lifetime, respectively), the final neutrino spectrum can be obtained. 

Since both $pp$ and $p\gamma$ processes produce neutrinos while other proton cooling processes suppress the final neutrino spectrum, the suppression factor taking into account various proton cooling processes is expressed as \citep{murase2008,wang2009,xiao2016} $\zeta_{p,{\rm sup}}(\epsilon_{\nu_i}) = (t'^{-1}_{p,pp} + t'^{-1}_{p,p\gamma}) / t'^{-1}_{p,{\rm cool}}$, where $t'^{-1}_{\rm cool} = t'^{-1}_{p,pp} + t'^{-1}_{p,p\gamma} + t'^{-1}_{p,\rm BH} + t'^{-1}_{p,\rm syn} + t'^{-1}_{p,\rm IC} + t'^{-1}_{p,\rm ada}$. Similarly, the suppression factor due to meson cooling can be written as $\zeta_{i,{\rm sup}}(\epsilon_{\nu_i}) =t'^{-1}_{i,{\rm dec}} / t'^{-1}_{i,{\rm cool}}$, where $t'^{-1}_{i,{\rm cool}} = t'^{-1}_{i,{\rm dec}} + t'^{-1}_{i,{\rm had}} + t'^{-1}_{i,{\rm syn}} + t'^{-1}_{i,{\rm IC}} + t'^{-1}_{i,{\rm ada}}$. One can obtain the neutrino spectrum in each neutrino production channel for a single event,
\begin{equation}
    \epsilon_{\nu_i}^2F_{\nu_i} = \frac{N_iE_{\rm iso}\zeta_{p,{\rm sup}}(\epsilon_{\nu_i})\zeta_{i,{\rm sup}}(\epsilon_{\nu_i})}{4\pi D_L^2\ln(\epsilon'_{p,{\rm max}} / \epsilon'_{p,{\rm min}})},
\end{equation}
where $N_\pi = N_{\mu_\pi} = 0.12$, $N_K = 0.009$, and $N_{\mu_K} = 0.003$, the neutrino energy is $\epsilon_{\nu_i} = a_i\Gamma_{\rm h}\epsilon'_p$ with $a_\pi = a_{\mu_\pi} = 0.05$, $a_K = 0.10$, and $a_{\mu_K} = 0.033$, where $D_{\rm L}$ is the luminosity distance, and $\ln(\epsilon'_{p,{\rm max}} / \epsilon'_{p,{\rm min}})$ is the normalized factor with $\epsilon'_{p,{\rm min}}\approx\Gamma_{\rm h}m_pc^2$. Highly efficient acceleration is assumed here (i.e., the acceleration efficiency $\zeta_p \simeq 1$) so that $\epsilon_{\rm acc} \approx \zeta_p (1 - \epsilon_e - \epsilon_B) \approx 0.8 \sim 1$ which is in accord with the fiducial value of the baryon loading parameter $\xi_{\rm acc}\simeq\epsilon_{\rm acc} / \epsilon_e \approx 10$ \citep{murase2007}. 

\begin{figure}[tb]
    \centering
    \includegraphics[width = 1\linewidth , trim = 50 34 70 57, clip]{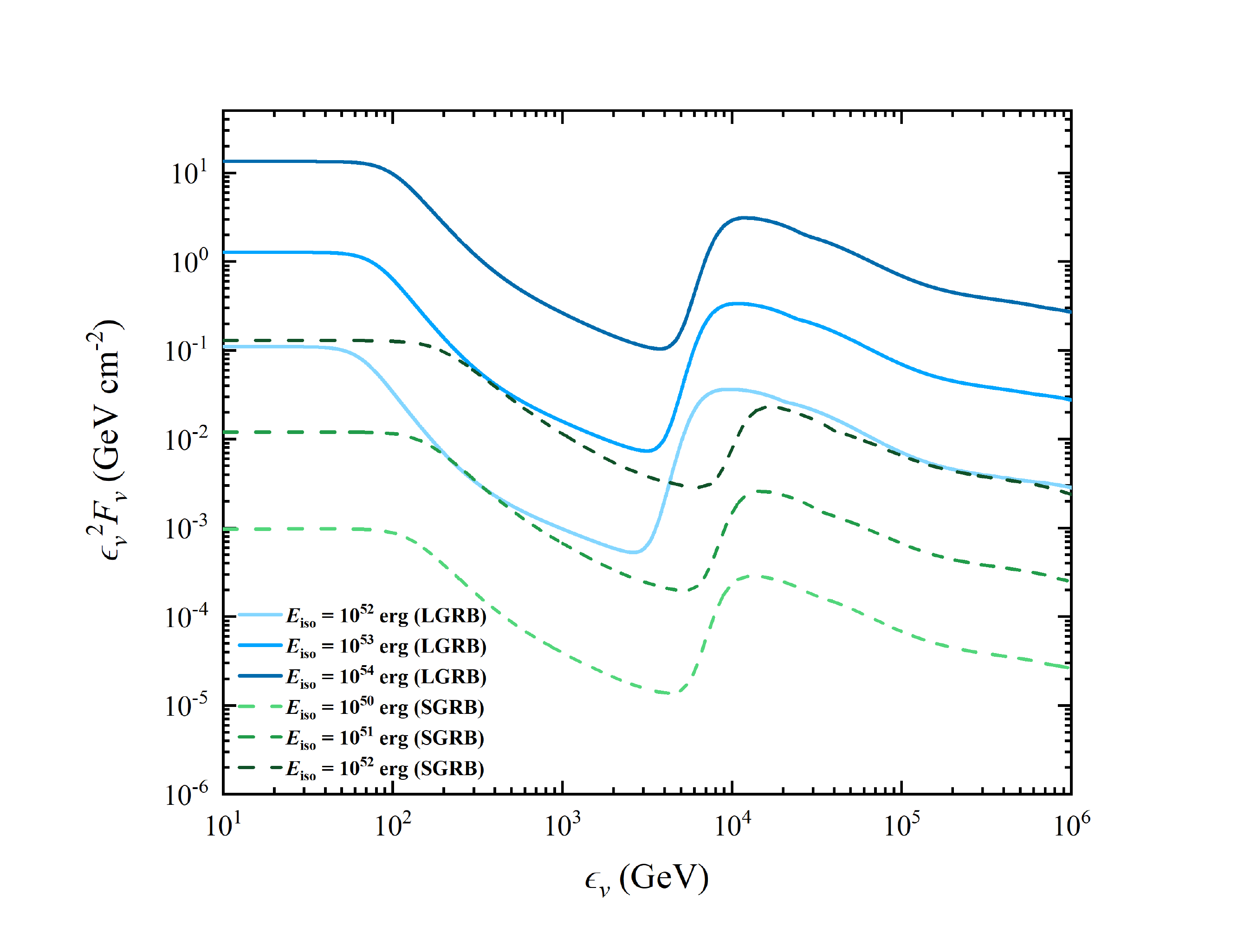}
    \caption{Expected all-flavor neutrino fluence as a function of neutrino energy $\epsilon_\nu$ for GRBs at $D_{\rm L} = 100\,{\rm Mpc}$. The three solid curves from light blue to dark blue are for choked LGRBs with three different isotropic jet energies: $E_{\rm iso} = 10^{52}$, $10^{53}$, and $10^{54}\,{\rm erg}$. The three dashed curves from light green to dark green are for choked SGRBs with three different isotropic jet energies: $E_{\rm iso} = 10^{50}$, $10^{51}$, and $10^{52}\,{\rm erg}$.}
    \label{fig:Spectrum}
\end{figure}

Figure\,\ref{fig:Spectrum} shows the all-flavor fluence of a single burst at $D_{\rm L} = 100\,{\rm Mpc}$. The fluence is mainly determined by the isotropic energy. The dip around a few TeV is caused by the suppression of neutrino production due to the Bethe-Heitler process. Low-energy neutrinos are dominated by $pp$ interactions, and the neutrino spectrum above $\sim1\,{\rm TeV}$ that we are interested in mainly attributes to $p\gamma$ interactions. Both $pp$ and $p\gamma$ processes are efficient as shown in Figure\,\ref{fig:Spectrum}, since both the $pp$ optical depth $n_p'\sigma_{pp}r_{\rm stall}/\Gamma_{\rm h}\sim40(25)$ and the $p\gamma$ optical depth  $n_\gamma'\sigma_{p\gamma}r_{\rm stall}/\Gamma_{\rm h}\sim10^6(2\times10^5)$ for a classical LGRB (SGRB) are quite large  \citep{murase2008} \cite[considering $\sigma_{pp}\approx5\times10^{-26}\,{\rm cm}^2$ and $\sigma_{p\gamma}\approx5\times10^{-28}\,{\rm cm}^2$][for rough estimations]{eidelman2004}. 

\begin{table}[tb]
    \footnotesize
    \centering
    \caption{Neutrino Bursts Detection}
    \begin{tabular}{cccc}
    \multicolumn{4}{c}{Number of detected $\nu_\mu$ from single event at $100\,{\rm Mpc}$} \\
    \hline
    \hline
    Model ($E_{\rm iso}/{\rm erg}$)  & IceCube (Up) & IceCube (Down) & Gen2(Up)\\
    \hline
    LGRB ($10^{52}$) & $0.16$ & $0.016$ & $0.76$\\
    LGRB ($10^{53}$) & $1.5$ & $0.15$ & $6.8$\\
    LGRB ($10^{54}$) & $13$ & $1.4$ & $61$\\
    SGRB ($10^{50}$) & $1.2\times10^{-3}$ & $1.4\times10^{-4}$ & $5.7\times10^{-3}$ \\
    SGRB ($10^{51}$) & $0.011$ & $1.3\times10^{-3}$ & $0.050$\\
    SGRB ($10^{52}$) & $0.094$ & $0.012$ & $0.44$\\
    \hline
    \hline
    \multicolumn{4}{c}{Joint GW + Neutrino Detection Rate} \\
    \hline
    \hline
    \multicolumn{2}{c}{Era(GW/Neutrino)} & \multicolumn{2}{c}{Detection Rate (${\rm yr}^{-1}$)} \\
    \hline
    \multicolumn{2}{c}{O4/IceCube} &  \multicolumn{2}{c}{$f_{\rm AGN}[0.006, 1.3]\times10^{-1}$} \\
    \multicolumn{2}{c}{O5/IceCube} &  \multicolumn{2}{c}{$f_{\rm AGN}[0.007, 1.3]\times10^{-1}$} \\
    \multicolumn{2}{c}{Voyager/Gen2} &  \multicolumn{2}{c}{$f_{\rm AGN}[0.041, 8.0]\times10^{-1}$} \\
    \multicolumn{2}{c}{ET/Gen2} &  \multicolumn{2}{c}{$f_{\rm AGN}[0.042, 8.8]\times10^{-1}$} \\
    \hline
    \hline
    \end{tabular}
    \label{tab:Detection}
\end{table}

\section{Neutrino Bursts Detection}

The expected number of muon neutrinos $\nu_\mu$ from an on-axis GRB event detectable by IceCube and IceCube-Gen2 (Gen2) can be calculated by
\begin{equation}
    N(\epsilon_\nu>1\,{\rm TeV}) = \int_{1\,{\rm TeV}}^{\epsilon_{\nu,{\rm max}}}d\epsilon_\nu F_\nu(\epsilon_\nu) A_{\rm eff}(\epsilon_\nu),
\end{equation}
where $A_{\rm eff}$ is the effective area of the detector. We obtain the effective areas of IceCube for the up- and down-going events from \cite{Aartsen2017}. The effective volume of the Gen2 is larger than that of IceCube by a factor of $\sim 10$, corresponding to a factor of $\sim10^{2/3}$ times larger in the effective area \citep{Aartsen2017}. The number of detected $\nu_\mu$ \cite[e.g.,][]{harrison2002} from a single event located at $100\,{\rm Mpc}$ are shown in Table\,\ref{tab:Detection} (after considering neutrino oscillation). If a classical LGRB occurs in an AGN disk at $100\,{\rm Mpc}$, we expect $\sim2\,(7)$ neutrino events from a single event detected by IceCube (Gen2). The neutrino flux from a SGRB is lower, and the detection for a single choked SGRB is possible only with Gen2 given that the SGRB has a high energy and occurs in the  Northern Hemisphere.

We simulate the joint GW+neutrino detection rate for neutron star mergers occurring in AGN disks. \cite{mckernan2020} showed that the local event rate densities $\dot{\rho}_0$ for BNS and NSBH mergers in the AGN channel are $\dot{\rho}_{0,\rm BNS}\sim f_{\rm AGN}[0.2, 400]\,{\rm Gpc}^{-3}\,{\rm yr}^{-1}$ and $\dot{\rho}_{0,\rm NSBH}\sim f_{\rm AGN}[10, 300]\,{\rm Gpc}^{-3}\,{\rm yr}^{-1}$, respectively, where $f_{\rm AGN}$ is the fraction of the observed BBH in the AGN channel. The redshift distribution $f(z)$ we adopt is a the model invoking a log-normal delay timescale distribution with respect to star formation history, as applied to the study of SGRBs \citep{wanderman2015,virgili2011,sun2015}. The GW horizon distances for BNS and NSBH mergers in different eras are obtained from \citep{abbott2018,maggiore2020,hild2011,zhu2020}. By considering the beaming correction factor $f_{\rm b}\approx(\theta_0 + 1/\Gamma_{\rm h})^2/2$ and assuming that all BNS and $20\%$ NSBH mergers \citep{mckernan2020} in AGN disks can power classical SGRBs, we show the results of joint GW+neutrino detection rates in Table\,\ref{tab:Detection}. Such joint detections appear difficult with the current GW and neutrino detectors, but could be possible in the future with next generation GW and neutrino detectors.

\begin{figure}[tb]
    \centering
    \includegraphics[width = 1\linewidth , trim = 45 34 70 57, clip]{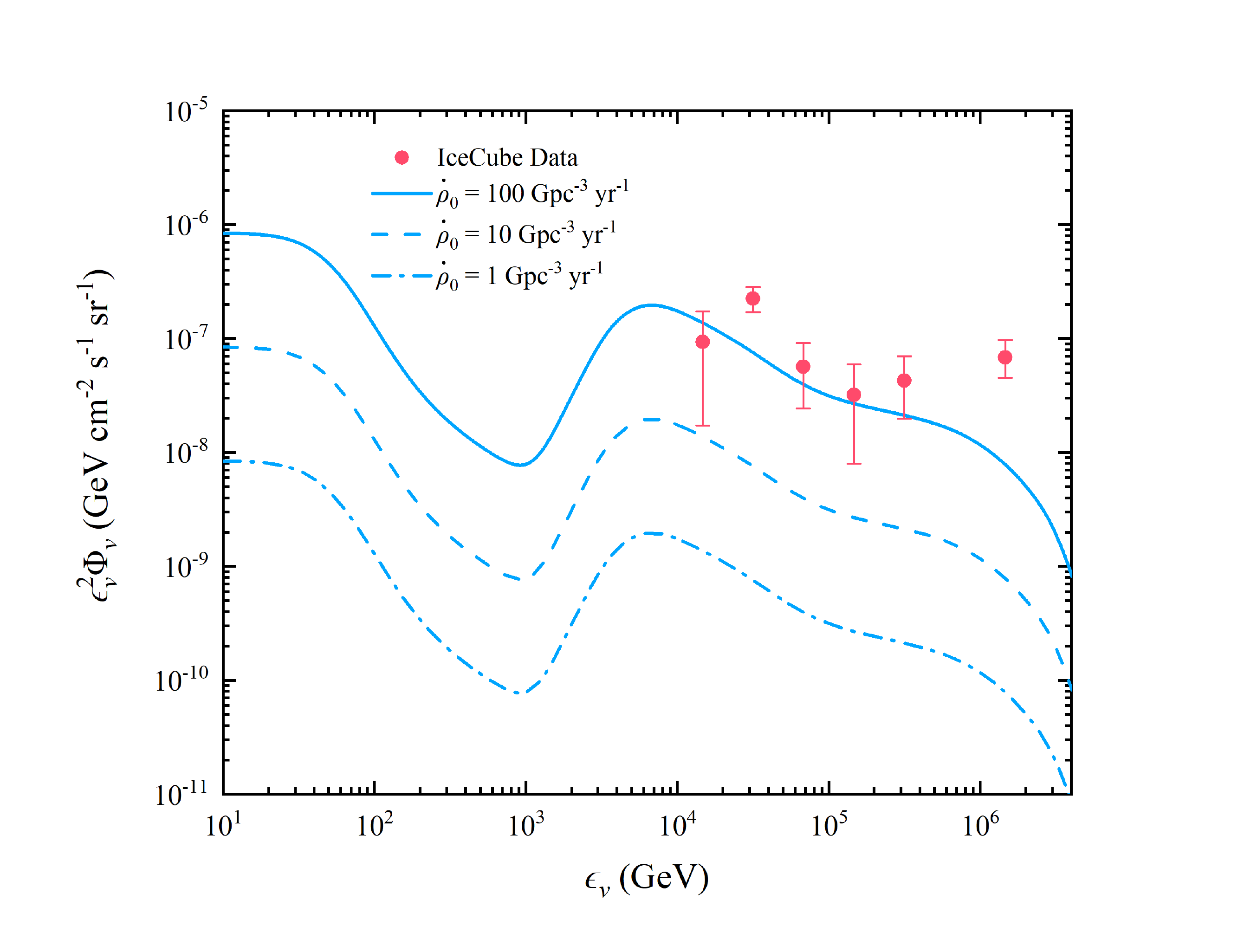}
    \caption{Expected all-flavor diffuse neutrino fluence contributed from choked LGRBs in AGN disks as a function of neutrino energy $\epsilon_\nu$. Three local event rates are considered: $100$ (solid line), $10$ (dashed line), and $1\,{\rm Gpc}^{-3}\,{\rm yr}^{-1}$ (dashed-dot line). The pink circles are observed diffuse neutrino fluence measured by IceCube \citep{aartsen2015a}.  }
    \label{fig:DiffusedSpectrum}
\end{figure}

\section{Neutrino Diffuse Emission}

The diffuse neutrino fluence can be estimated as \cite[e.g.,][]{razzaque2004}
\begin{equation}
    \epsilon_{\nu,{\rm obs}}^2\Phi_{\nu,{\rm obs}} = \epsilon_{\nu,{\rm obs}}^2 f_{\rm b} \int_0^{z_{\rm max}}dz{\dot{\rho}_0f(z)}F_{\nu}(\epsilon_{\nu,{\rm obs}})\frac{dV}{dz},
\end{equation}
where $\epsilon_{\nu,{\rm obs}} = \epsilon_\nu / (1 + z)$, and $dV/dz = 4\pi D_{\rm L}^2c/(1+z)|dt/dz|$ is the comoving volume element with $(dt/dz)^{-1} = - H_0(1 + z)\sqrt{\Omega_\Lambda + \Omega_{\rm m}(1+z)^3}$. The standard $\Lambda$CDM cosmology with $H_0 = 67.8\,{\rm km}\,{\rm s}^{-1}\,{\rm Mpc}^{-1}$, $\Omega_{\Lambda} = 0.692$, and $\Omega_{\rm m} = 0.308$ \citep{planck2016} is applied.

In view that the energy of a typical SGRB is much smaller than that of a LGRB and that their event rate density may not be much greater than that of long GRBs, the contribution of choked SGRBs in AGN disks to the neutrino background should be much lower than that of choked LGRBs. We thus only consider the contribution from the latter. Since the cosmic evolution of AGN and star formation rate is not significant \cite[e.g,][]{madau2014}, we assume that LGRBs in AGN disks are classical LGRBs which closely track the star formation history. We adopt the $f(z)$ distribution based on \cite{yuksel2008}. 

We show in Figure\,\ref{fig:DiffusedSpectrum} the diffuse neutrino fluence by considering three values of the local event rate density due to choked LGRBs in AGN disks are poorly constrained \cite[since these events do not show up as classical GRBs][]{zhu2021,perna2021}. In an extreme case, if $\dot{\rho_0} = 100\,{\rm Gpc}^{-3}\,{\rm yr}^{-1}$, i.e. comparable to that of low-luminosity GRBs \cite[$\sim 164^{+98}_{-65}\,{\rm Gpc}^{-3}\,{\rm yr}^{-1}$,][]{sun2015}, most of the observed neutrino background fluence could be interpreted by choked LGRBs in AGN disks. If $\dot{\rho_0} = 10\,{\rm Gpc}^{-3}\,{\rm yr}^{-1}$, which is $\sim5\%$ that of low-luminosity GRBs, the neutrinos from such events can contribute up to $\sim10\%$ of the observed diffuse neutrino background fluence. Future observations of shock breakout transients from AGN disks \citep{zhu2021,perna2021} will better constrain the cosmological event rate density of these choked GRBs in the AGN channel, leading to a better estimation of their contribution to the neutrino background.

\section{Discussion}

Choked LGRBs and SGRBs in AGN disks are ideal targets for multi-messenger observations. Besides neutrino emission discussed in this paper, they can also produce electromagnetic signals from optical to $\gamma$-ray bands \citep{zhu2021,perna2021}. For choked LGRBs, associated SNe could be directly discovered by time-domain survey searches \citep{assef2018}. Within several years of operation by IceCube and Gen2, neutrino bursts from single events would be possible to be directly detected. Choked SGRBs in AGN disks are emitters of electromagnetic, neutrino and GW signals. Future joint observations of electromagnetic, neutrino, and GW signals can reveal the existence these hitherto speculated transient population in AGN disks, shedding light into the interplay between AGN accretion history and the star formation and compact binary merger history in the universe.

Besides classical GRBs from core collapse of massive stars, neutron star mergers and binary BH mergers \citep[e.g.,][]{bartos2017,kaaz2021,kimura2021}, accretion of a single BH \citep{wang2021} and accretion-induced collapse of NSs \citep{Perna2021b} embedded within AGN disks were also studied recently. Such jets driven by embedded AGN objects could potentially be choked as well and hence produce high-energy neutrinos.

The stalling radius for the jets choked in the AGN disk materials is $\sim 10^{12}-10^{13}\,{\rm cm}$. On the other hand, in the traditional GRB model, this radius is also where $\gamma$-ray emission is generated (e.g. via internal shocks). However, before the jet breaks out from the star, the jet Lorentz factor is smaller so that internal shock radius would be further in closer to the central engine. Furthermore, as shown by \cite{perna2021}, in a large parameter space, an external shock into the disk material develops before internal shocks. Even if internal shocks form, $\gamma$-ray photons generated in these shocks have a small mean free path due to huge Thompson optical depth of the disk $(\tau_{\rm T} \approx \rho_0 \sigma_{\rm T} H / m_p \approx 4\times 10 ^3 \rho_{0,-10}H_{14})$. A small fractional of $\gamma$-ray photons may be consumed via $p\gamma$ production to produce neutrinos, but most $\gamma$-ray photons would be trapped and degraded in energy before escaping the disk. 
Nonetheless, the cocoon shock breakout from the AGN disk can potentially produce low-luminosity $\gamma$-ray emission \citep{zhu2021} which has similar observed properties as low-luminosity GRBs. This would happen only if the disk environment at the location where the GRB occurs is less dense while the choked GRB jet is powerful enough. However, $\gamma$-ray photons from low-luminosity GRBs are relatively soft \citep[$\lesssim100\,{\rm keV}$; e.g.,][]{campana2006,soderberg2006,nakar2015} which are below the energy coverage range of Fermi-LAT \citep{ackermann2015}. Thus, choked GRB jets in AGN disks do not significantly contribute to the isotropic $\gamma$-ray background.

\acknowledgments

We thank an anonymous referee for thoughtful suggestions. We thank Zhuo Li, Di Xiao, Ming-Yang Zhuang, Yuan-Qi Liu, and Di Zhang for valuable comments. The work of J.P.Z is partially supported by the National Science Foundation of China under Grant No. 11721303 and the National Basic Research Program of China under grant No. 2014CB845800. K.W is supported by the National Natural Science Foundation under grants 12003007 and the Fundamental Research Funds for the Central Universities (No. 2020kfyXJJS039). Y.P.Y is supported by National Natural Science Foundation of China grant No. 12003028 and Yunnan University grant No.C176220100087. Y.W.Y is supported by the National Natural Science Foundation of China under Grant No. 11822302, 11833003. H.G. is supported by the National Natural Science Foundation of China under Grant No. 11722324, 11690024, 11633001, the Strategic Priority Research Program of the Chinese Academy of Sciences, Grant No. XDB23040100 and the Fundamental Research Funds for the Central Universities.

\bibliography{NeutrinoProduction}{}
\bibliographystyle{aasjournal}

\end{document}